\documentclass[12pt]{article}
\usepackage{graphicx}
\newcounter{subequations}
\title{Magnetic Field Dependence of $T_{AB}$ in Superfluid $^3He$}

\author{G.A. Baramidze and G.A. Kharadze\\
\\
{\it Andronikashvili Institute of Physics, Tamarashvili str. 6,}\\
{\it 0177 Tbilisi, Georgia (e-mail: gogi@iphac.ge)}}

\begin{document}
\maketitle

\begin{abstract}
The magnetic field dependence of the $A\rightarrow B$ transition
temperature $T_{AB}$ in the superfluid $^3He$ is reconsidered in
order to take into account the linear-in-field contribution beyond
the approximation used in Ref.[7]. In the high field region, where
the quadratic-in-field contribution prevails, the well known
answer is restored. On the other hand,it is pointed out that the
Fermi liquid effects shift the observability of the
linear-in-field region to rather low magnetic fields.
\end{abstract}

Soon after the discovery of the superfluidity of liquid $^3He$ it
was established that near critical temperature $T_c(P)$  the phase
diagram of superfluid state experiences a profound modification
under the action of even small external magnetic fields [1, 2, 3].
This modification shows up in elimination of the direct normal to
$B$ phase transition over entire phase diagram in favour of an
anisotropic $A$ phase.

The $A\rightarrow B$ transition temperature $T_{AB}(H)<T_c$ can be
established by equating the free energies $F_A(H)$ and $F_B(H)$.
Near the zero-field transition temperature $T_c(P)$ the free
energy can be found by minimizing the Ginzburg-Landau functional
\begin{equation}
F_S=3\alpha_{\mu\nu}<\Delta_{\mu}\Delta_{\nu}^{\ast}>
+F_S^{(4)}(\vec\Delta),
\end{equation}
where the fourth order contribution in the order-parameter
$\vec{\Delta}(\hat{k})$ reads as

\begin{eqnarray}
F_S^{(4)}=9\{\beta_1<\Delta_{\mu}\Delta_{\mu}><\Delta_{\nu}^{\ast}\Delta_{\nu}^{\ast}>
+\beta_2<|\vec{\Delta}|^2>^2+
\beta_3<\Delta_{\mu}\Delta_{\nu}><\Delta_{\mu}^{\ast}\Delta_{\nu}^{\ast}>\nonumber\\
\\
+\beta_4<\Delta_{\mu}\Delta_{\nu}^{\ast}><\Delta_{\mu}^{\ast}\Delta_{\nu}>
+\beta_5<\Delta_{\mu}\Delta_{\nu}^{\ast}><\Delta_{\mu}\Delta_{\nu}^{\ast}>\}.\nonumber
\end{eqnarray}

In Eqs. (1) and (2) the spin-vector $\vec{\Delta}$  is defined
according to the general expression for the order-parameter
$A_{\mu i}$ of the spin-triplet $p$-wave Cooper condensate:
$\Delta_{\mu}(\hat{k}) =A_{\mu i}\hat{k}_i$, and the angular
brackets $<\cdots>$ stand for an average over the momentum
direction $\hat{k}$ on the Fermi surface. In presence of a
magnetic field $\vec{H}$ the tensor coefficient $\alpha_{\mu\nu}$
of the second order term in Eq.(1) reads as

\begin{equation}
\alpha_{\mu\nu}=\alpha\delta_{\mu\nu}+
ig_1\varepsilon_{\mu\nu\lambda}H_{\lambda}+ g_2H_{\mu}H_{\nu},
\end{equation}
where
\renewcommand{\theequation}{\arabic{equation}\alph{subequations}}
\setcounter{subequations}{1}
\begin{equation}
\alpha=\frac{1}{3}N_F\ln\biggl(\frac{T}{T_c}\biggl),
\end{equation}
\addtocounter{equation}{-1}
\addtocounter{subequations}{1}
\begin{equation}
g_1H=-\frac{1}{3}N_F\eta h,
\end{equation}
\addtocounter{equation}{-1}
\addtocounter{subequations}{1}
\begin{equation}
g_2H^2=\frac{1}{3}N_F\kappa h^2,
\end{equation}
\renewcommand{\theequation}{\arabic{equation}}
with $h=\hbar\gamma H/(2k_B T_c)=H/H_o$. The dimensionless
coefficients $\eta$ and $\kappa$ could be considered as the
phenomenological quantities although their values can be estimated
according to the microscopic calculations. In the weak-coupling
approximation [4]
\begin{equation}
\eta_{wc}=\frac{N_F^{\prime}}{N_F}k_B T_c
\ln\biggl(\frac{2\gamma_E\omega_c}{\pi T_c} \biggl)
\end{equation}
where $N_F^{\prime}$  stands for the derivative of the
quasiparticle DOS with respect to the energy at the Fermi level. A
detailed calculations which take into account the linear-in-field
corrections to the Fermi liquid parameters are performed in Ref.
[5].

As to the  $\kappa$-coefficient, it stems from the free energy
part
\begin{equation}
\delta F_H^{(2)}=\frac{1}{2}\delta\chi_s H^2,~~~~~~\delta\chi_s=
\chi_s-\chi_N,
\end{equation}
where $\chi_s(\chi_N)$ stands for the magnetic susceptibility of
the superfluid (normal) phase. For the $B$ phase near $T_c$,
$\delta F_H^{(2)}=g_2 H^2\Delta^2$ where $g_2 H^2$ is given
according to Eq. (4c) with
\begin{equation}
\kappa=\frac{7\zeta(3)}{4\pi^2}\frac{1}{(1+F_o^a)^2}
\end{equation}
Here the Fermi liquid parameter $F_o^a\cong-3/4$ and is weakly
pressure dependent.

As it is well known, the linear-in-field term (4b) is the origin
of a tiny splitting of the $A$ phase (due to a small asymmetry of
the density of quasiparticle states at the Fermi level). The
quadratic-in-field contribution (4c) gives rise to the magnetic
field suppression of $\Delta_{\uparrow\downarrow}$ component of
the energy gap of the $B$ phase.

Based on the argument that the term (4b) is rather small, in the
majority of considerations of  $T_{AB}=T_{AB}(H)$ this term is
usually discarded, and in a standard way one starts from the
following expressions for the equilibrium free energies of the $A$
and $B$ phases:
\renewcommand{\theequation}{\arabic{equation}\alph{subequations}}
\setcounter{subequations}{1}
\begin{equation}
F_A=-\frac{1}{4\beta_{245}}\alpha^2,
\end{equation}
\addtocounter{equation}{-1}
\addtocounter{subequations}{1}
\begin{equation}
F_B=-\frac{1}{2(3\beta_{12}+\beta_{345})}
\biggl[\frac{3}{2}\alpha^2+g_2H^2\alpha+\frac{2\beta_{12}+\beta_{345}}{2\beta_{345}}
(g_2 H^2)^2\biggl].
\end{equation}
\renewcommand{\theequation}{\arabic{equation}}
Since here the linear-in-field contribution (proportional to
$g_1$) is dropped, the action of the magnetic field appears only
in $F_B$.

Equating (8a) and (8b) it is found that $T_{AB}(H)$ is to be
obtained from the equation
\begin{equation}
\varphi=\alpha^2+a\alpha+b=0
\end{equation}
with
\begin{eqnarray}
a=2P_1g_2H^2,\nonumber\\
\\
b=P_1^2(1-q_1)(g_2H^2)^2\nonumber
\end{eqnarray}
where the coefficients $P_1$ and $q_1$ are defined in the
Appendix. From Eq.(9) it follows the answer for
$\tau_{AB}=1-T_{AB}/T_c$:
\begin{equation}
\tau_{AB}=P_1(1+\sqrt{q_1})\kappa h^2
\end{equation}
which reproduces a well known result obtained in Ref.[6].

Now we turn to the role of the linear-in-field contribution to
$T_{AB}(H)$. This question was first posed in Ref.[7]. The
starting point is the construction of the expressions for
$F_{A2}(H)$ and $F_B(H)$, the contribution (4b) being taken into
account. In a standard way it is established that
\begin{equation}
F_{A2}=-\frac{1}{4\beta_{245}}\biggl[\alpha^2-
\frac{\beta_{245}}{\beta_5}g_1^2H^2\biggl],
\end{equation}

\begin{eqnarray}
F_B&=&-\frac{1}{2(3\beta_{12}+\beta_{345})}\Biggl\{\frac{3}{2}\alpha^2+g_2H^2\alpha
+\frac{3\beta_{12}+\beta_{345}}{\beta_4-(3\beta_1+\beta_{35})}g_1^2H^2+\nonumber\\
\\
&+&\frac{2\beta_{12}+\beta_{345}}{2\beta_{345}}(g_2H^2)^2+\nonumber\\
\nonumber\\
&+&\frac{(3\beta_{12}+\beta_{345})^2}{(3\beta_1+\beta_{35}-\beta_4)^2}
\biggl[\frac{\beta_1}{\beta_{345}}\frac{g_1^2g_2H^4}{\alpha}+
\frac{\beta_1(3\beta_{12}+\beta_{345})}{4(3\beta_1+\beta_{35}-\beta_4)^2}
\frac{g_1^4H^4}{\alpha^2}\biggl]\Biggl\}.\nonumber
\end{eqnarray}

Comparison of $F_{A2}$ and $F_B$ gives an equation for
$T_{AB}(H)$:

\begin{equation}
\varphi(T_{AB}/T_c)=\alpha^4+a\alpha^3+b\alpha^2+c\alpha+d=0
\end{equation}
where now
\begin{eqnarray}
a&=&2P_1g_2H^2,\nonumber\\
\nonumber\\
b&=&-P_1P_2P_3g^2_1H^2+P_1^2(1-q_1)(g_2H^2)^2,\nonumber\\
\\
c&=&-P_1P_2^2P_4g_1^2g_2H^4,\nonumber\\
\nonumber\\
d&=&-P_1P_2^3P_5(g_1H)^4.\nonumber
\end{eqnarray}
The definition of all the coefficients $P_a$ is given in the
Appendix.

In Ref.[7] the second and third lines in Eq.(13), containing terms
on the order $H^4$ were neglected. This has the following
influence on the answer for $T_{AB}(H)$: i) neglection of the
contribution proportional to $(g_2H^2)^2$ makes it impossible one
to reproduce the correct answer given by Eq.(11) (the term
$\sqrt{q_1}$ will be lost),  ii) neglection of the contribution
collected in the square brackets of the third line of Eq.(13)
changes the answer for the linear-in-field contribution to
$T_{AB}(H)$. To avoid this drawbacks we address Eq.(14) and, as a
first step, perform the variable transformation $\alpha\rightarrow
x-\frac{1}{4}a$, after which an equation for $x(T_{AB}/T_c)$ is
obtained:
\begin{equation}
\varphi(T_{AB}/T_c)=x^4+Ax^2+Bx+C=0
\end{equation}
with the coefficients
\begin{eqnarray}
A&=&-\biggl[P_1P_2P_3(g_1H)^2+\frac{1}{2}P_1^2(1+2q_1)(g_2H^2)^2\biggl],\nonumber\\
\nonumber\\
B&=&q_1P_1^3(g_2H^2)^3+q_2P_1P_2g_1^2g_2H^4,\nonumber\\
\\
C&=&-P_1P_2^3P_5(g_1H)^4+\frac{1}{16}P_1^4(1-4q_1)(g_2H^2)^4+\nonumber\\
\nonumber\\
&+&\frac{1}{4}P_1^2P_2(P_1P_3-2q_2)(g_1g_2H^3)^2.\nonumber
\end{eqnarray}

It is to be noticed, that as a result of the variable
transformation used, in Eq. (16) the cubic term is absent. At the
same time the coefficient of the linear term $B$ is zero in the
weak-coupling approximation.

In order to solve Eq. (16) we use the decomposition
$x=x_o-\sqrt{q_1}P_1g_2H^2$ which generates the decomposition
$\varphi=\varphi_o+\delta\varphi$ with $\varphi_o$ being the
solution of the equation

\begin{equation}
\varphi_o=x_o^4 +A_ox^2_o+C_o=0,
\end{equation}
where $A_o$ and $C_o$ are the coefficients $A$ and $C$ taken at
$q_1=q_2=0$.

From Eq. (18) it is found that

\begin{equation}
x_o(T_{AB}/T_c)=-\sqrt{P_o^2(g_1H)^2+\frac{1}{4}P_1^2(g_2H^2)^2}
\end{equation}
where

\begin{equation}
P_o^2=\frac{1}{2}P_1P_2P_3\bigl[1+\sqrt{1+4P_2P_5/P_1P_3^2}\bigl].
\end{equation}

The direct inspection shows that $\delta\varphi$ is the sum of
terms proportional to the powers of $q$ and the minimal power in
$H$ contained in $\delta\varphi$ is $H^5$. For this reason we can
use an approximation with $\delta\varphi$ disregarded, and as a
result it is found that
\begin{eqnarray}
\alpha(T_{AB}/T_c)=x-\frac{1}{4}a=x_o-\sqrt{q_1}P_1g_2H^2
-\frac{1}{2}P_1g_2H^2=\nonumber\\
\\
=-P_1\biggl(\frac{1}{2}+\sqrt{q_1}\biggl)g_2H^2-
\sqrt{P_o^2(g_1H)^2+\frac{1}{4}P_1^2(g_2H^2)^2}.\nonumber
\end{eqnarray}

Using this result we finally have the following simple answer for
$\tau_{AB}$:
\begin{equation}
\tau_{AB}=P_1\biggl(\frac{1}{2}+\sqrt{q_1}\biggl)\kappa h^2+
\sqrt{P_o^2(\eta h)^2+\frac{1}{4}P_1^2(\kappa h^2)^2}.
\end{equation}

By introducing a characteristic magnetic field
$H_{\ast}=2(P_o/P_1)(\eta/\kappa)H_o$ the asymptotic regions where
the quadratic-in-field $(H>>H_{\ast})$ and the linear-in-field
$(H<<H_{\ast})$ contributions to $T_{AB}$ prevail are found:
\begin{equation}
\tau_{AB}=\Biggl\{
\begin{array}{cc}
P_1(1+\sqrt{q_1})\kappa h^2, & H>>H_{\ast} \\
P_o\eta h, & H<<H_{\ast}.\\
\end{array}
\end{equation}

For $H>>H^{\ast}$ the well known result [4,8] is reproduced.

In order to isolate the role of the linear-in-field contribution
to $T_{AB}$ it is convenient to consider
\begin{equation}
\delta\tau_{AB}(h)=\tau_{AB}(h)-(1+\sqrt{q_1})P_1\kappa h^2,
\end{equation}
and construct graphically
\begin{equation}
f(h)=\delta\tau_{AB}(h)/h=P_o\eta(\sqrt{1+(h/h_{\ast})^2}-h/h_{\ast}),
\end{equation}
where the scaling value $h_{\ast}=2(P_o/P_1)(\eta/\kappa)$.

Below $f(h)$ is plotted for the pressure $P=10~bar~(P_o\eta\simeq
5,4\cdot 10^{-2}, h_{\ast}\simeq 1,6\cdot 10^{-2})$.

\begin{figure}[t]
\includegraphics[width=0.8 \linewidth]{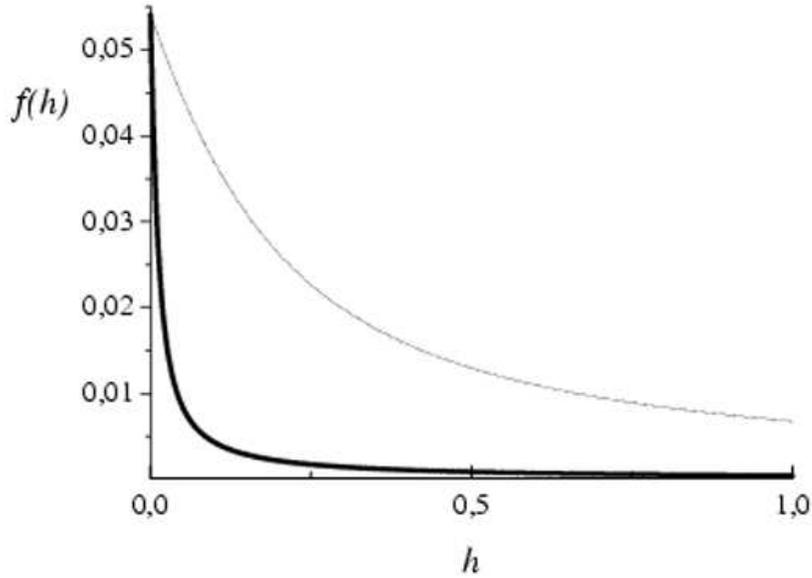}
\caption{\label{fig:pop} The bold curve corresponds to the Landau
parameter $F_o^a=-3/4$ and the thin curve is constructed at
$F_o^a=0$. It is obvious that the Fermi liquid effect shifts the
linear-in-field contribution to rather small values of
$H<<H_{\ast}=h_{\ast}H_o\simeq 420~G$.}
\end{figure}

\vspace{1cm} {\bf Acknowledgements}

The authors highly appreciate the valuable discussions with late
Dr. Sh.Nikolaishvili.

\vspace{2cm} {\bf Appendix} \vspace{1cm}

The coefficients $P_a~(a=1,2,3,4,5)$ and $q_a~(a=1,2)$ introduced
in the main text are defined as follows:
\begin{eqnarray}
P_1&=&\frac{\beta_{245}}{2\beta_{345}-3\beta_{13}},\nonumber\\
\nonumber\\
P_2&=&\frac{3\beta_{12}+\beta_{345}}{\beta_4-3\beta_1-\beta_{35}},\nonumber\\
\nonumber\\
P_3&=&\frac{\beta_{45}-3\beta_1-\beta_3}{-\beta_5},\nonumber\\
\nonumber\\
P_4&=&\frac{-2\beta_1}{\beta_{345}},\nonumber\\
\nonumber\\
P_5&=&\frac{-\beta_1}{2(\beta_4-3\beta_1-\beta_{35})}.\nonumber
\end{eqnarray}

\begin{eqnarray}
q_1&=&\frac{(3\beta_{12}+\beta_{345})(2\beta_{13}-\beta_{345})}{\beta_{245}\beta_{345}},\nonumber\\
\nonumber\\
q_2&=&P_1P_3-P_2P_4,\nonumber
\end{eqnarray}
where $\beta_{ij\cdots}=\beta_i+\beta_j+\cdots$.

The coefficients $q_1$ and $q_2$ contain only the strong-coupling
corrections $\delta\beta_i$ defined according to the decomposition
\begin{eqnarray}
\beta_1=-\beta_o+\delta\beta_1,\nonumber\\
\nonumber\\
\beta_2=2\beta_o+\delta\beta_2,\nonumber\\
\nonumber\\
\beta_3=2\beta_o+\delta\beta_3,\nonumber\\
\nonumber\\
\beta_4=2\beta_o+\delta\beta_4,\nonumber\\
\nonumber\\
\beta_5=-2\beta_o+\delta\beta_5,\nonumber\\
\nonumber\\
\beta_o=\frac{7\zeta(3)}{240}\frac{N_F}{(\pi
T_c)^2}.\nonumber
\end{eqnarray}


\begin{thebibliography}{8}

\bibitem   .W.J. Gully et al., {\it Phys. Rev.} {\bf A 8}, 1633 (1973).

\bibitem   .D.N. Paulson et al., {\it Phys. Rev. Lett.} {\bf 32}, 1098 (1974).

\bibitem   .J.D. Feder et al., {\it Phys. Rev. Lett.} {\bf 47}, 428 (1981).

\bibitem   .V. Ambegaonar and D. Mermin, {\it Phys. Rev. Lett.}
{\bf 30}, 81 (1973).

\bibitem   .K. Bedell and K. Quadar, {\it Phys. Rev.} {\bf B 30},
2894 (1984).

\bibitem   .A.L. Fetter, in {\it "Quantum Statistics and the
Many-Body Problem"}, Plenum, NY, 1975; {\it J.Low Temp. Phys.}
{\bf 23}, 245 (1976).

\bibitem   .K. Levin and O. Walls, {\it Phys. Rev.} {\bf B15},
4256 (1977).

\bibitem   .Y.H. Tang et al., {\it Phys. Rev. Lett.}
{\bf 67}, 1775 (1991).
\end{thebibliography}
\end{document}